 \documentclass[final,authoryear,times,twocolumn]{elsarticle}
\makeatletter
\def\ps@pprintTitle{%
 \let\@oddhead\@empty
 \let\@evenhead\@empty
 \def\@oddfoot{}%
 \let\@evenfoot\@oddfoot}
\makeatother

\usepackage{graphicx}
\usepackage{url}
\usepackage{amssymb}

\journal{New Astronomy}

\begin{document}

\begin{frontmatter}



\title{Astrophysical data mining with GPU. A case study: genetic classification of globular clusters}


\author[ref1,ref3]{S. Cavuoti}
\author[ref2]{M. Garofalo}
\author[ref3,ref1]{M. Brescia\corref{cor1}}
\ead{brescia@oacn.inaf.it}
\author[ref1]{M. Paolillo}
\author[ref2]{A. Pescape'}
\author[ref1,ref3,ref4]{G. Longo}
\author[ref2]{G. Ventre}

\cortext[cor1]{Corresponding Author}

\address[ref1]{Department of Physics, University Federico II, Via Cinthia 6, I-80126 Napoli, Italy}
\address[ref2]{Department of Computer Engineering and Systems, University Federico II, Via Claudio 21, I-80125 Napoli, Italy}
\address[ref3]{INAF, Astronomical Observatory of Capodimonte, Via Moiariello 16, I-80131 Napoli, Italy}
\address[ref4]{Visiting Associate, California Institute of Technology, Pasadena, CA 91125, USA}

\begin{abstract}
We present a multi-purpose genetic algorithm, designed and implemented with GPGPU / CUDA parallel computing technology. The model was derived from our CPU serial implementation, named GAME (Genetic Algorithm Model Experiment). It was successfully tested and validated on the detection of candidate Globular Clusters in deep, wide-field, single band HST images.
The GPU version of GAME will be made available to the community by integrating it into the web application DAMEWARE (DAta Mining Web Application REsource,  \url{http://dame.dsf.unina.it/beta_info.html}), a public data mining service specialized on massive astrophysical data. Since genetic algorithms are inherently parallel, the GPGPU computing paradigm leads to a speedup of a factor of $200\times$ in the training phase with respect to the CPU based version.
\end{abstract}


\end{frontmatter}


\section{Introduction}
\label{Intro}

In the present data intensive era, when almost all sciences are blessed by an avalanche of high quality, complex data, computing has assumed a new methodological significance \citep{hey2009}.
In fact, the extraction of the useful information (such as patterns, trends, etc), from such datasets cannot be performed with traditional methods and requires, as support to the decisional process, the usage of automatic or semiautomatic methods and procedures \citep{banerji2010, djorgovski2012}.
However, the sheer size (e.g., billions of galaxies and/or thousand of features) is such that also the most traditional Data Mining (DM) methods need to be at least revisited, in order to properly scale to the multi-Tera or even Petabyte regime \citep{fabbiano2010}.
For what scalability is concerned,  there are good reasons to believe that in the near future most aspects of computing will see exponential growth in bandwidth, but sub-linear or no improvements at all in latency. Moore's Law \citep{moore1965} will continue to deliver exponential increases in memory size but the speed with which data can be transferred between memory and Central Processing Units (CPUs) will remain more or less constant and marginal improvements can only be made through advances in
caching technology \citep{akhter2006}.
Certainly Moore's law still leaves room for the creation of parallel computing capabilities on single chips by packing multiple CPU cores onto it, but the clock speed that determines the speed of computation is constrained to remain limited by a thermal wall \citep{sutter2005}.
Thus it has to be expected that individual CPUs will not become much faster in the near future, even though they will have the parallel computing power and storage capacity that today can be provided only via specialized hardware. From the application development point of view, to build programs capable to exploit low latency
in multi core CPUs will require a fundamental shift from the currently sequential or parallel programming, to a mixture of parallel and distributed programming.
In recent years a few groups have began to test the possibility to solve some of the above problems by means of  General Purpose Graphical Processing Unit (known as GPGPU; \citealt{nvidia2012}) parallel technologies. Among the most known parallel programming models, we quote the OpenCL\footnote{\url{http://www.khronos.org/registry/cl/sdk/1.0/docs/man/xhtml/}} (Open Computing Language), which is a foundation layer of platform-independent tools; CUDA\footnote{\url{https://developer.nvidia.com/what-cuda}} (Compute Unified Device Architecture): a general purpose architecture that leverages the parallel compute engine within GPUs to solve many complex computational problems in a more efficient way than on a single CPU and, finally, OpenACC\footnote{\url{http://www.openacc-standard.org/}}, a set of compiling directives allowing programmers to create high-level CPU+GPU applications without the need to explicitly manage the parallel architecture configuration and its communication with the CPU host.
In astronomy, pioneering attempts to use the Graphical Processing Unit (GPU) technology have been made in an handful of applications. GPUs were attempted mainly to speed up numerical simulations (e.g. \cite{nakasato2011, valdez2012}) of various types from traditional N-body to fluid dynamics; but also to perform complex repetitive numerical tasks such as computation of correlation functions \citep{ponce2012}, real time processing of data streams \citep{barsdell2012}, and complex visualization \citep{hassan2012}, to quote just a few.
In this paper we present the implementation on GPUs of the multi-purpose Genetic Algorithm Model Experiment (GAME), capable to deal with both regression and classification problems. The serial version of this algorithm was implemented by some of the authors to be deployed on the DAta Mining \& Exploration (DAME) hybrid distributed infrastructure and made available to the community through the DAMEWARE (DAME Web Application REsource; \cite{brescia2010, djorgovski2012}) web application.
The present work is structured as follows: Sec. \ref{game} presents the design and serial implementation of the GAME model; Sec. \ref{Astrodata} describes the knowledge base (astrophysical dataset), used for the experiments; Sec. \ref{gpuimpl} is dedicated to the overview of the GPU technology in the astrophysical context, while the experiments and results are described in Sec. \ref{experiment}, just before our conclusions.

\section{GAME}\label{game}
Genetic algorithms \citep{mitchell1998} are derived from Darwin's evolution law \citep{darwin1859} and are intrinsically parallel in their learning evolution rule and processing data patterns. This implies that the parallel computing paradigm can lead to a strong optimization in terms of processing performances.
Genetic and evolutionary algorithms  are an important family of supervised Machine Learning models which tries to emulate the evolutionary laws that  allow living organisms to adapt their life style to the outdoor environment and survive.\\
In the optimization problems which can be effectively tackled with genetic evolutionary models, a key concept is the evaluation function (or \emph{fitness} in evolutionary jargon), used to determine which possible solutions get passed on to multiply and mutate into the next generation of solutions. This is usually done by analyzing the \emph{genes}, which refer to
some useful information about a particular solution to the problem under investigation.
The fitness function looks at the genes, performs some qualitative assessment and then returns the fitness value for that solution. The remaining parts of the genetic algorithm discard all solutions having a poor fitness value, and accept only those with a good fitness value. Frequently, the fitness function has not a closed mathematical expression and, for instance, in some cases it may be given under the form of the simulation of a real physical system.\\
GAME is a pure genetic algorithm specially designed to solve supervised optimization problems related with regression and classification functionalities.  It is scalable to efficiently manage massive datasets and is based on the usual genetic evolution methods which will be shortly outlined in what follows.\\
As a supervised machine learning technique, GAME requires to select and create the data parameter space, i.e. to create the data input patterns to be evaluated. It is important in this phase to build homogeneous patterns, i.e. with each pattern having the same type and number of parameters (or features), as well as to prepare the datasets which are needed for the different experiment steps: training, validation and test sets, by splitting the parameter space into variable subsets to be submitted at each phase. The dataset must include also target values for each input pattern, i.e. the desired output values, coming from any available knowledge source.\\
From an analytic point of view, in the data mining domain, a pattern is a single sample of a real problem, composed by a problem-dependent number of features (sometimes called attributes), which is characterized by a certain S/N (Signal-to-Noise) ratio. In other words, such pattern contains an unspecified amount of information, hidden among its correlated features, that an unknown analytical relation is able to transform into the correct output. Usually, in real cases (such as the astronomical ones), the S/N ratio is low, which means that feature correlation of a pattern is \emph{masked} by a considerable amount of noise (intrinsic to the phenomenon and/or due to the acquisition system); but the unknown correlation function can ever be approximated with a polynomial expansion.

The generic function of a polynomial sequence is based on these simple considerations:\\
Given a generic dataset with N features and a target $t$, let $pat$ be a generic input pattern of the dataset, $pat = (f_1, ..., f_N, t)$ and $g(x)$ a generic real function.
The representation of a generic feature $f_i$ of a generic pattern, with a polynomial sequence of degree $d$ is:

\begin{equation}
G({f_i}) \cong a_0 + a_1g({f_i}) + ... + a_dg^d(f_i )
\label{eq1}
\end{equation}

Hence, the k-th pattern $(pat_k)$ with N features may be written as:

\begin{equation}
Out({pat_k}) \cong \sum\limits_{i = 1}^N {G({f_i}) \cong a_0 + \sum\limits_{i = 1}^N {\sum\limits_{j = 1}^d {a_j g^j({f_i})} } }
\label{eq2}
\end{equation}

and the target $t_k$, related to pattern $pat_k$, can be used to evaluate the approximation error (the fitness) of the input pattern to the expected value:

\begin{equation}
E_k  = ({t_k - Out({pat_k})})^2
\label{eq3}
\end{equation}

If we generalize eq.~\ref{eq2} to an entire dataset, with NP number of patterns $(k = 1,..., NP)$, the \emph{forward} phase of the GA (Genetic Algorithm) consists of the calculation of NP expressions of the eq.~\ref{eq2}, which represent the polynomial approximation of the dataset.\\
In order to evaluate the fitness of the NP patterns as extension of eq.~\ref{eq3}, the Mean Square Error (MSE) or Root Mean Square Error (RMSE) may be used:

\begin{equation}
MSE  = \frac{\sum\limits_{k = 1}^{NP} {(t_k - Out(pat_k))^2 }} { NP }
\label{eq4}
\end{equation}

\begin{equation}
RMSE  = \sqrt{\frac{\sum\limits_{k = 1}^{NP} {(t_k - Out(pat_k))^2 }} { NP }}
\label{eq5}
\end{equation}\\

On the basis of the above, we obtain a GA with the following characteristics:

\begin{itemize}\addtolength{\itemsep}{-0.5\baselineskip}
\item The expression in eq.~\ref{eq2} is the \emph{fitness} function;
\item The array $(a_0,..., a_M)$ defines M genes of the generic chromosome (initially they are generated random and normalized between -1 and +1);
\item All the chromosomes have the same size (constrain from a classic GA);
\item The expression in eq.~\ref{eq3} gives the standard error to evaluate the fitness level of the chromosomes;
\item The population (genome) is composed by a number of chromosomes imposed from the choice of the function $g(x)$ of the polynomial sequence.
\end{itemize}

The number of chromosomes is determined by the following expression:

\begin{equation}
NUM_{chromosomes}  = (B \cdot N) + 1
\label{eq6}
\end{equation}

where $N$ is the number of features of the patterns and $B$ is a multiplicative factor that depends on the $g(x)$ function, which in the simplest case is just 1, but can be even 3 or 4 in more complex cases. The parameter $B$ also influences the dimension of each chromosome (number of genes):

\begin{equation}
NUM_{genes}  = (B \cdot d) + 1
\label{eq7}
\end{equation}

where $d$ is the degree of the polynomial. In more general terms, the polynomial expansion within the fitness function, defined in eq.~\ref{eq2}, could be arbitrarily chosen and we conceived the implementation code accordingly. In the specific context of this work we decided to adopt the trigonometric polynomial expansion, given by the following expression (hereinafter \emph{polytrigo}):

\begin{equation}
g(x) = a_0  + \sum\limits_{m = 1}^d {a_m \cos (mx) + } \sum\limits_{m = 1}^d {b_m \sin (mx)}
\label{eq8}
\end{equation}

In order to have $NP$ patterns composed by $N$ features, the expression using eq.~\ref{eq2} with degree $d$, is:

\begin{eqnarray}\label{eq9}\nonumber
Out(pat_{k = 1...NP} ) \cong \sum_{i = 1}^{N} G(f_i ) \cong a_0  + \\
+ \sum_{i = 1}^{N} \sum_{j = 1}^{d} a_j \cos (jf_i ) +
\sum_{i = 1}^{N}
\sum_{j= 1}^d b_j \sin (jf_i )
\end{eqnarray}

In the last expression we have two groups of coefficients (sine and cosine), so $B$ will assume the value $2$.
Hence, the generic genome (i.e. the population at a generic evolution stage), will be composed by $2N+1$ chromosomes, given by eq.~\ref{eq6}, each one with $2d+1$ genes $[a_0, a_1,...,a_d, b_1, b_2,...,b_d]$, given by eq.~\ref{eq7}, with each single gene (coefficient of the polynomial) in the range $[-1, +1]$ and initially randomly generated.\\

By evaluating the goodness of a solution through the MSE or RMSE metrics, it may happen that a better solution in terms of MSE corresponds to a worse solution for the model, for example when we have a simple crispy
(jargon for classification with sharp boundaries between adjacent classes) classification problem with two patterns (class types 0 and 1).\\
As an example, if the solutions are, respectively, 0.49 for the class 0 and 0.51 for the class 1, the efficiency (i.e. the percentage of objects correctly classified) is 100\%, with a $MSE = 0.24$. But a solution of 0 for the class 0 and 0.49 for the class 1 (efficiency of 50\%), gives back a $MSE = 0.13$ and  consequently the model will prefer the second solution, although with a lower efficiency.

In order to circumvent this problem, we decided to implement in GAME what we could call a \emph{convergence tube} or \emph{Threshold MSE} (TMSE):\\

\begin{equation}
\left\{ \begin{array}{l}
 if ( {t_k  - Out( {pat_k } )} )^2  > R \Rightarrow E_k  = ( {t_k  - Out( {pat_k } )} )^2  \\
 if ( {t_k  - Out( {pat_k } )} )^2  \le R \Rightarrow E_k  = 0 \\
 \end{array} \right.
\label{eq10}
\end{equation}\\

where $R$ is a user defined numerical threshold in the $]0,1[$ range.\\
In the previous example, by using $R = 0.5$ in eq.~\ref{eq10}, we obtain a $TMSE = 0$ and $TMSE = 0.13$, respectively in the first and second cases, thus indicating that the first solution is better than the second one, as expected.\\
Through several cycles, the population of chromosomes, originally started as a random generation (typically following a normal distribution), is replaced at each step by a new one which, by having higher fitness value, represents a step forward towards the best population (i.e. the optimal solution).
Typical representations used are the binary code or the real values in [-1, 1] for each element (gene) of a chromosome. \\

As it is typical for the supervised machine learning models, the classification experiment with GAME is organized in a sequence of functional use cases:

\begin{itemize}\addtolength{\itemsep}{-0.5\baselineskip}
\item \emph{train}: the first mandatory case, consisting into submitting training datasets in order to build and store the best GA population, where best is in terms of its problem solving capability;
\item \emph{test}: case to be used in order to verify and validate the learning performance of the trained GA;
\item \emph{run}: the normal execution mode after training and validation;
\item \emph{full}: a workflow case, including \emph{train} and \emph{test} cases in sequence.
\end{itemize}

{The training phase is the most important use case, and is based on a user defined loop of iterations (schematically shown in Fig.~\ref{figure1}), whose main goal is to find the population of chromosomes (i.e. the solutions to the classification problem), having the maximum fitness.\\
Three types of random generation criteria can be adopted:
\begin{itemize}\addtolength{\itemsep}{-0.5\baselineskip}
\item pseudo-random values from [-1, +1];
\item random values from the normal distribution in [-1, +1];
\item pseudo-random values from [0, 1].
\end{itemize}

\begin{figure}
\centering
\includegraphics[width=7cm,height=8.5cm]{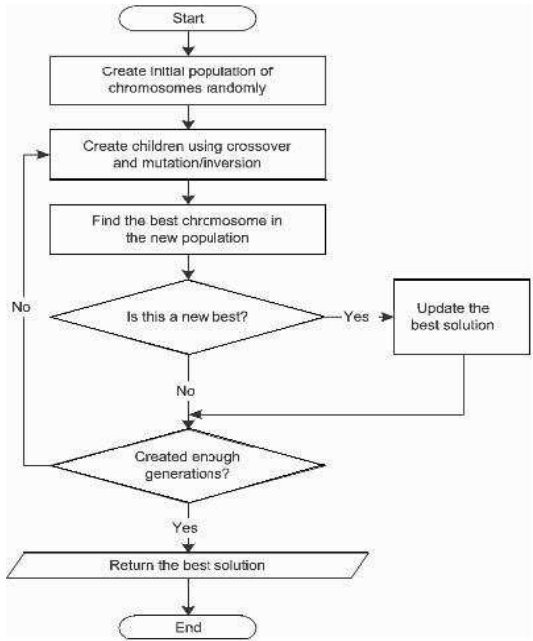}
\caption{The flow-chart of the CPU-based GAME model implementation.}
\label{figure1}
\end{figure}

At each training step there are the following options, which can be chosen by the user during the configuration phase:

\begin{itemize}\addtolength{\itemsep}{-0.5\baselineskip}
\item \emph{error function} type: MSE, TMSE or RMSE;
\item \emph{selecting criterion}: the genetic selection function to evolve the population of chromosomes at each training iteration. It is possible to choose between the classical \emph{ranking} and \emph{roulette} types \citep{mitchell1998};
\item \emph{Elitism}, which is a further mechanism to evolve the population. The user has to choose the number of copies of the winner chromosome, to be maintained unchanged in the next generation. Main role of elitism is indeed to preserve the best fitness along evolution \citep{mitchell1998}.
\end{itemize}

In the present work, the idea was to build a GA able to solve supervised crispy classification and regression problems, typically related to a high-complexity parameter space where the background analytic function is not known, except for a limited number of couples of input-target values, representing a sample of known solutions for a physical category of phenomena (also indicated as the Knowledge Base, or KB). The goal is indeed to find the best chromosomes so that the related polynomial expansion is able to approximate the correct classification of the input patterns. Therefore in our case the fitness function is evaluated in terms of the training error, obtained as the absolute difference between the target value and the polynomial expansion output for all patterns.

\section{The data}\label{Astrodata}

As mentioned in the introduction, the algorithm was tested on a specific astrophysical problem which can be traced back to the well known star/galaxy  (resolved/unresolved) separation.
Wide field multi-band photometry is usually required to identify GCs (Globular Clusters) in external galaxies, since such objects appear as unresolved sources in ground-based astronomical images and are hardly distinguishable from background galaxies.

Therefore, on ground based data, GCs are traditionally selected using methods based on colors and luminosities. This, however, introduces biases which can be partly circumvented using higher angular resolution images obtained from space facilities (i.e. Hubble Space Telescope, HST). These higher resolution images can be used also to measure GC properties such as sizes and structural parameters (core radius, concentration, etc.).

However, suitable HST data are challenging in terms of observing time since the optimal datasets would require to be: i) multi-band, to effectively select GC based on colors; ii) deep, in order to sample the majority of the GC population and ensure the high S/N required to measure structural parameters, see \cite{carlson2001} and iii) with wide-field coverage, in order to minimize projection effects as well as to study the overall properties of the GC populations, which often differ from those inferred from observations of the central region of a galaxy only.

It is therefore quite obvious that, to use single-band HST data would be much more effective in terms of observing costs. But such approach requires to carefully select the candidate GCs based on the available photometric and morphological parameters in order to avoid introducing biases in the final sample. In \cite{brescia2012a} some of us showed that the use of properly tuned data mining algorithms (in that case genetic algorithms), on top of a high efficient computing architecture, can yield very complete datasets with low contamination, even with single band photometry, thus minimizing the observing time requirements and allowing to extend such studies to larger areas and to the outskirts of nearby galaxies.

The algorithm used in that work, however, turned out to be rather expensive in terms of computing time and this triggered our interest in demonstrating that GPUs can be quite effective to speed up this types of computations.\\

The dataset used in our experiment consists in wide field HST observations of the giant elliptical NGC1399 in the Fornax cluster \citep{paolillo2011}.
This galaxy represents an ideal test case since: i) due to its distance (20 Mpc), a large fraction of its GC system can be covered with a limited number of observations; ii) GCs are only marginally resolved even by HST, thus allowing to verify our model in a \emph{worst-case} scenario.

The optical data were taken with the HST Advanced Camera for Surveys (ACS, program GO-10129), in the F606W filter, with an integration time of 2108 seconds for each field.
The observations were arranged in a 3x3 ACS mosaic, and combined into a single image using the MultiDrizzle routine \citep{koekemoer2002}.
The final scale of the images is 0.03 arcsec/pix, thus providing Nyquist sampling of the ACS Point Spread Function (PSF).\\
In this experiment we used two ancillary multiwavelength datasets: archival HST g-z observations, which cover the very central region of the galaxy ($\sim 10\%$ of the extension putatively covered by the GC system), and ground based photometry \citep{paolillo2011}. The latter is only available for 14\% of our sources, and due to background light contamination, is very incomplete close to the galaxy center.
In total 2740 sources of the catalog have multi-band photometry. Finally, the subsample of sources used to build our Knowledge Base necessary to train the GAME model is composed of the 2100 sources with all photometric and morphological information. These features are:

\begin{itemize}\addtolength{\itemsep}{-0.5\baselineskip}
\item The isophotal magnitude (feature 1); 	
\item Three aperture magnitudes (features 2--4) obtained through  circular apertures of radii 2, 6, and 20 arcsec respectively;
\item The Kron radius, the ellepticity and the FWHM of the image (features 5-7);
\item The structural parameters	(features 8-11) which are, respectively, the central surface brightness, the	core radius, the effective radius and the tidal radius.
\end{itemize}

The details of the procedure have been described in \citep{brescia2012a} therefore we shall just highlight a few, relevant points.

The traditional approach to GCs candidate selection would be to adopt the magnitude and color cuts reported in Table \ref{col_sel}, and highlighted in Figure \ref{col_mag} with a dashed line; the magnitude limit $z<22.5$ does not exploit the full depth of the HST data but it is adopted in order to be consistent with the $T1<23$ limit used for the ground-based colors, thus ensuring a uniform limit across the whole field-of-view. In the following, we assume that bona-fide GCs are represented by such sources, in order to explore how well different selection methods based on single band photometry are able to retrieve the correct population of objects.
In total our classification dataset consisted of 2100 patterns, each composed by 11 features (plus the target, corresponding to the classes GC or not GC used during the supervised training phase).

\begin{table}\centering\small
\begin{tabular}{lcc}
\hline
 & color cut & magnitude cut\\
\hline
Ground-based data & $1.0\leq C\!-\!T1<2.2$ & $T1<23$\\
HST data & $1.3\leq g\!-\!z<2.5$ & $z<22.5$ \\
\hline
\end{tabular}
\caption{Photometric selection criteria for GC candidates}
\label{col_sel}
\end{table}

\begin{figure*}\centering
\includegraphics[height=6.8cm,width=6.8cm]{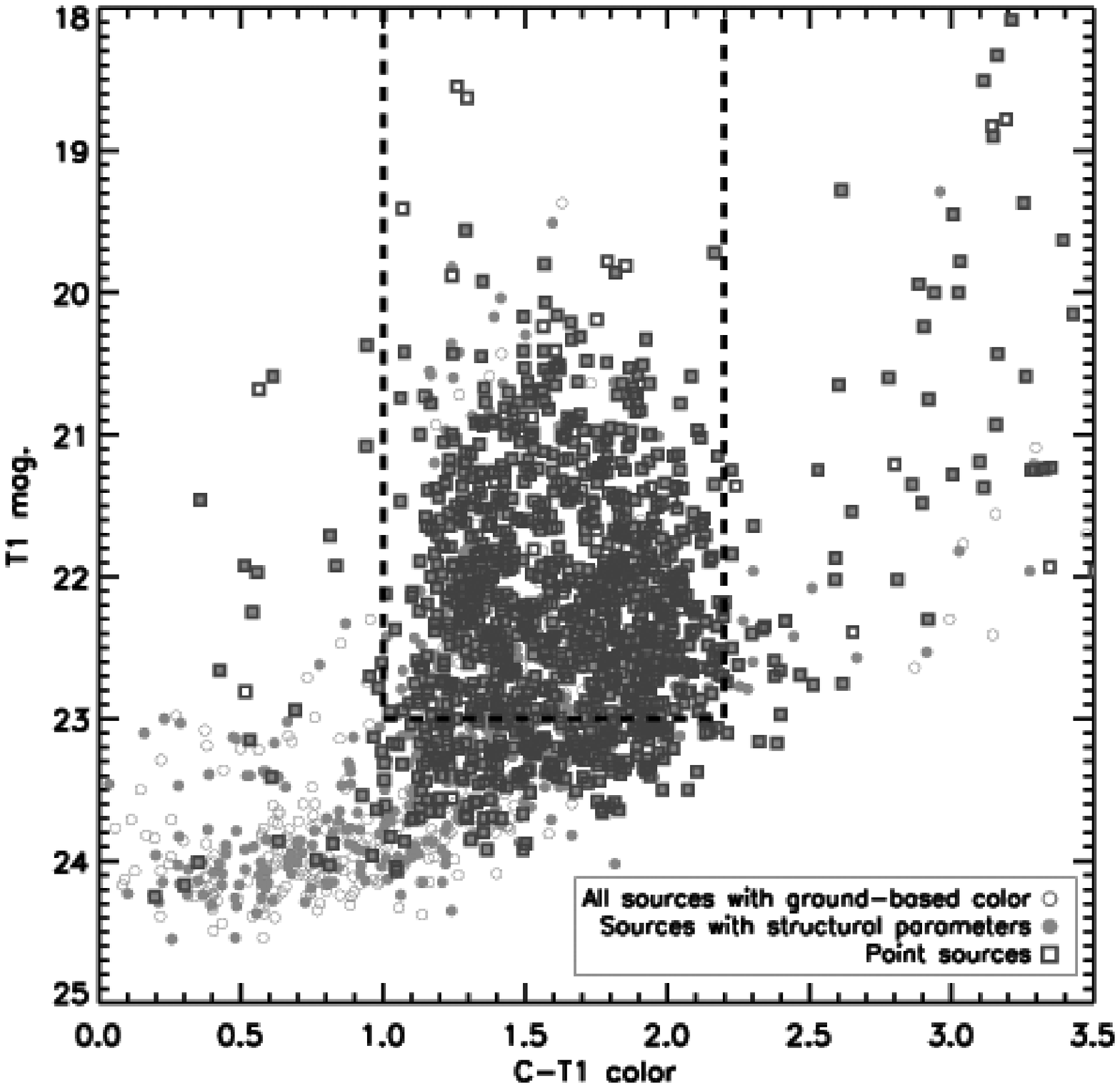}
\includegraphics[height=6.8cm,width=6.8cm]{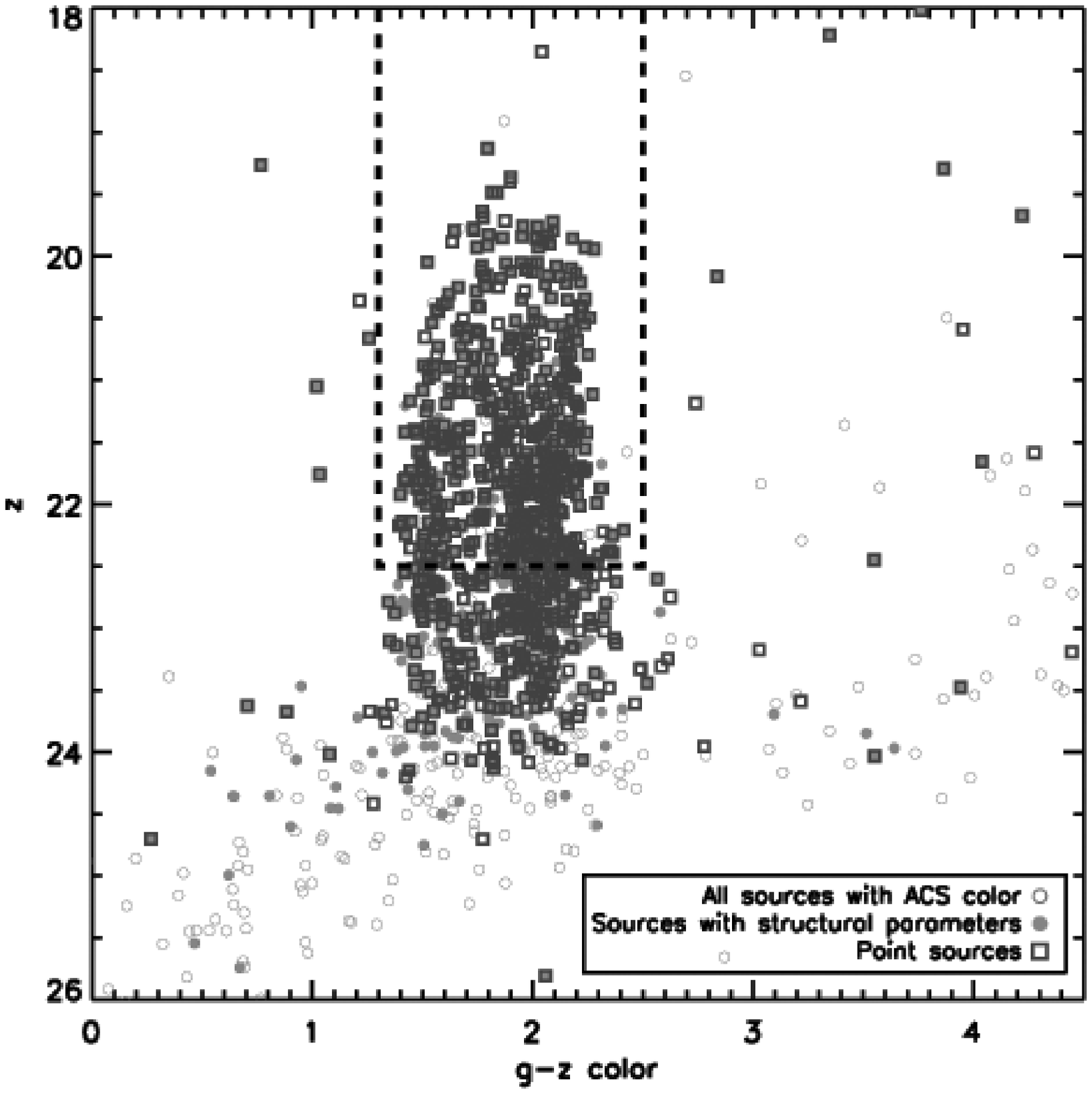}
\caption{Color-magnitude diagrams using $C\!-\!T1$ ground-based (left panel) and $g\!-\!z$ HST photometry (right panel). Ground-based photometry covers the whole FOV of our ACS mosaic, while HST colors are limited to the central ACS field. Open grey dots represent all sources in color catalogs while solid ones refer to the subsample with both color and structural parameters that represents our Knowledge Base. Blue squares mark point-like sources, i.e. sources with stellarity index $>0.9$, while the dashed line highlights the parameter space (Table \ref{col_sel}) used to select bona-fide GC.}
\label{col_mag}
\end{figure*}

\section{The GPU implementation of GAME}
\label{gpuimpl}

GPGPU stands for General Purpose Graphics Processing Units and was invented by M. Harris \citep{harris2003}, who first understood that GPU technology could be used also for other than graphic applications.
In general, graphic chips, due to their intrinsic nature of multi-core processors (many-core, when systems have more that 32 cores) and being based on hundreds of floating-point specialized processing units, allow many algorithms to achieve higher (one or two orders of magnitude) performance than usual CPUs.
Since 2009, the throughput peak ratio between GPU (many-core) and CPU (multi-core) was about $10:1$.
It must be noted that such values are referred mainly to the theoretical speed supported by such chips, i.e. 1 TeraFLOPS (where FLOPS stands for Floating point operations per second), against 100 GigaFLOPS. Such a large difference has pushed many developers to shift more computing-expensive parts of their programs on the GPUs. However, before 2006, GPUs were rather difficult to use, mainly because in order to access the graphic devices, programmers were conditioned to use specific APIs (Application Programming Interfaces), based on methods available through libraries like OpenGL\footnote{\url{http://www.opengl.org}} and Direct3D\footnote{\url{http://msdn.microsoft.com}}.
The limits and poor documentation of these APIs were strongly undermining the design and development of applications.\\
In order to better address the GPU-based design, starting from the CPU-based serial implementation of the GAME model, we chose the \emph{ad hoc} software development methodology \emph{APOD} (Assess, Parallelize, Optimize, and Deploy), see \citet{nvidia2012}.
The APOD design mechanism consists first in quickly identifying the portions of code that
can easily exploit and benefit from GPU acceleration, then in exploiting the optimization
task as fast as possible, see Fig.~\ref{figure2}. The
APOD is a cyclical process: initial speedups can be achieved, tested, and deployed quickly,
so that the cycle can start over to identify further optimization opportunities.
At each APOD cycle the identification of the parts of the code responsible for most of the execution time
is done by investigating the parallelized GPU accelerations.
An upper limit to performance improvement can be derived considering requirements and constraints,
and by applying the Amdahl \citep{amdahl1967} and Gustafson \citep{gustafson1988} laws.

\begin{figure}
\centering
\includegraphics[width=7cm]{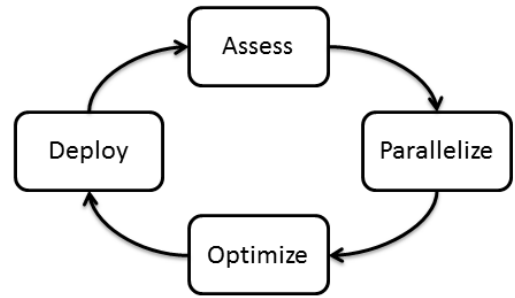}
\caption{The APOD parallel programming cyclic design mechanism.}
\label{figure2}
\end{figure}

By exposing the parallelism to improve performance and simply maintaining the code of sequential applications, we are able to ensure also the maximum parallel throughput on GPU platform. This could be as simple as adding a few preprocessor directives, such as OpenMP and OpenACC, or it can be done by calling an existing GPU-optimized library such as cuBLAS, cuFFT, or Thrust. Specifically, Thrust \citep{bell2011} is a parallel C++ template library resembling the C++ STL (Standard Template Library), described in \citet{stepanov1995}, which provides a rich collection of data parallel primitives such as scan, sort, and reduce, that can be composed together to implement complex algorithms with concise, readable source code.
After the parallelization step is complete, we can optimize the code to improve the performance. This phase is based on (i) maximizing parallel execution, identifying the most computationally expensive parts of the code, (ii) optimizing memory usage and (iii) minimizing low bandwidth host-to-device data transfers.\\

\subsection{The parallelization of GAME}

In all execution modes (use cases), GAME exploits the \emph{polytrigo} function defined by eq.~\ref{eq8}, consisting in a polynomial expansion in terms of sum of sines and cosines.
In particular, this calculation involves the following operations:
\begin{enumerate}[1.]
\item randomly generate the initial population of chromosomes;
\item calculate the fitness functions to find and sort the best chromosomes in the population;
\item evaluate the stop criteria (error threshold or maximum number of iterations);
\item stop or use the genetic operators to evolve the population and go to 2.
\end{enumerate}

Specifically in the training use case, the \emph{polytrigo} is used at each iteration to evaluate the fitness for all chromosomes of the population. It is indeed one of the critical aspects to be investigated in the parallelization process.\\

\begin{figure*}
\centering
\includegraphics[width=12cm]{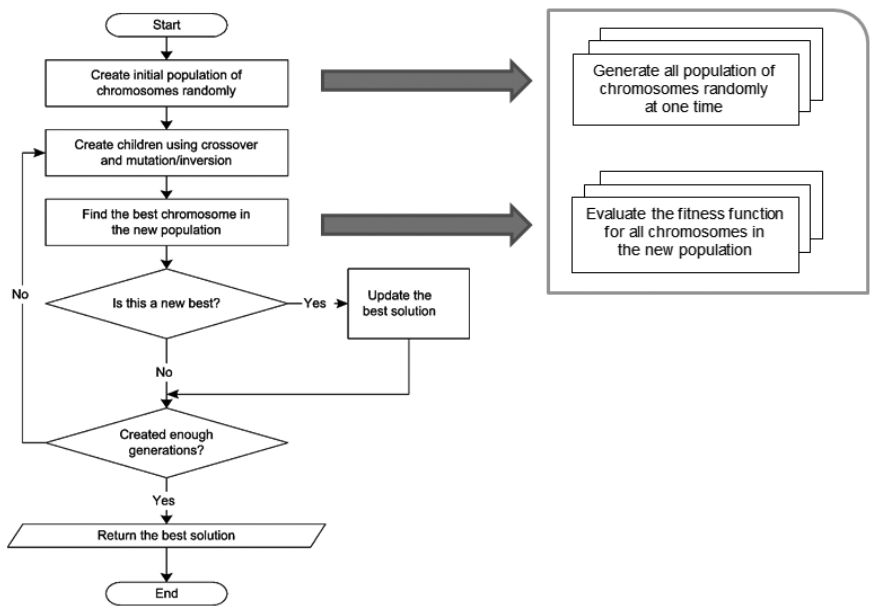}
\caption{The flow-chart of the GPU-based GAME model implementation.}
\label{figure3}
\end{figure*}

The main concern in designing the software architecture for the GPUs is to analyse the partition of work: i.e. which part of the work should be done on the CPU rather than on the GPU.
As shown in Fig.~\ref{figure3}, coherently with the APOD process we have analyzed and identified the time consuming critical parts to be parallelized by executing them on the GPU. They are the generation of random chromosomes and the calculation of the fitness function of chromosomes.
The key principle is that we need to perform the same instruction simultaneously on as much data as possible. By adding the number of chromosomes to be randomly generated in the initial population as well as during each generation, the total number of involved elements is never extremely large but it may occur with a high frequency. This is because also during the population evolution loop a variable number of chromosomes are randomly generated to replace older individuals. To overcome this problem we may generate a large number of chromosomes randomly \emph{una tantum}, by using them whenever required. On the contrary, the evaluation of fitness functions involves all the input data, which is assumed to be massive datasets, so it already has an intrinsic data-parallelism. The function \emph{polytrigo} takes about $\sim75\%$ of the total execution time of the application, while the total including child functions amounts to about $\sim87.5\%$ of total time execution. This indeed has been our first candidate for parallelization.\\
In order to give a practical example, for the interested reader, we report the source code portions related to the different implementation of the \emph{polytrigo} function, of the serial and parallelized cases.

\medskip

\noindent
{\it C++ serial code for \emph{polytrigo} function (eq.~\ref{eq8}):}
\begingroup
\fontsize{7pt}{7pt}
\begin{verbatim}
for (int i = 0; i < num_features; i++) {
   for (int j = 1; j <= poly_degree; j++) {
      ret +=  v[j] * cos(j * input[i]) + v[j + poly_degree] *
              * sin(j * input[i]); } }
\end{verbatim}
\endgroup

\medskip

\noindent
{\it CUDA C (Thrust) parallelized code for \emph{polytrigo} function (eq.~\ref{eq8}):}
\begingroup
\fontsize{7pt}{7pt}
\begin{verbatim}
struct sinFunctor {  __host__ __device__
 double operator()(tuple <double, double> t) {
    return sin(get < 0 > (t) * get < 1 > (t)); }};
struct cosFunctor {  __host__ __device__
 double operator()(tuple <double, double> t) {
    return cos(get < 0 > (t) * get < 1 > (t)); }};
thrust::transform(thrust::make_zip_iterator(
    thrust::make_tuple(j.begin(), input.begin())),
    thrust::make_zip_iterator(
       thrust::make_tuple(j.end(), input.end())),
       ret.begin(), sinFunctor(), cosFunctor());
\end{verbatim}
\endgroup

In the serial code the vector $v[]$ is continuously updated in the inner loop by means of the polynomial degree, while the vector $input[]$ (i.e. the elements of the input dataset) is used in each calculation but never altered. Therefore, in the parallelized version we rewrite the function by calculating in advance the sums of sines and cosines, storing the results in two vectors that are used in the function  \emph{polytrigo} at each iteration. This brings huge benefits because we calculate the time consuming trigonometric functions only once rather than at every iteration, so exploiting the parallelism on large amount of data.\\
From the time complexity point of view, by assuming to have as many GPU cores as population chromosomes, the above CUDA C code portion would take a constant time, rather than  the polynomial time required by the corresponding C++ serial code.\\
The part of the code which remains serialized runs on the host unless
an excessive host-to-device transfer slows down the process, in which case
also the serial part of the code runs on the device.
Having analyzed the application profile, we apply either Amdahl or Gustafson law to estimate an upper limit of the achievable speedup.
Once we have located a hotspot in our application's profile assessment, we used Thrust library to expose the parallelism in that portion of our code as a call to an external function. We then executed this external function onto the GPU and retrieved the results without requiring major changes to the rest of the application.\\
We adopted the Thrust code optimization in order to gain, at the cost of lower speedup, a rapid development and a better code readability. There are three high-level optimization techniques that we employed to yield significant performance speedups when using Thrust:

\begin{enumerate}[1.]
\item Fusion: In computations with low arithmetic intensity, the ratio of calculations per memory access are constrained by the available memory bandwidth and do not fully exploits the GPU. One technique for increasing the computational intensity of an algorithm is to fuse multiple pipeline stages together into a single one. In Thrust a better approach is to fuse the functions into a single operation $g(f(x))$ and halve the number of memory transactions. Unless f and g are computationally expensive operations, the fused implementation will run approximately twice as fast as the first approach. Fusing a transformation with other algorithms is a worthwhile optimization. Thrust provides a transform iterator which allows transformations to be fused with any algorithm;
\item Structure of Arrays (SoA): An alternative way to improve memory efficiency is to ensure that all memory accesses benefit from coalescing, since coalesced memory access patterns are considerably faster than non-coalesced transactions. The most common violation of the memory coalescing rules arises when using an Array of Structures (AoS) data layout. An alternative to the AoS layout is the SoA approach, where the components of each structure are stored in separate arrays. The advantage of the SoA method is that regular access to its components of a given vector is fusible;
\item Implicit Sequences: the use of implicit ranges, i.e. those ranges whose values are defined programmatically and not stored anywhere in memory. Thrust provides a counting iterator, which acts like an explicit range of values, but does not carry any overhead. Specifically, when counting iterator is de-referenced it generates the appropriate value \emph{on the fly} and yields that value to the caller.
\end{enumerate}

\section{The Experiment}
\label{experiment}

As already mentioned, the classification dataset consists of 2100 rows (input patterns) and 12 columns (features), 11 as input and the latter class target (class labels, respectively, 0 for not GC and 1 for GC objects), as described in Sec. \ref{Astrodata}. The performance was evaluated on several hardware platforms. We compared our production GPU code with a CPU implementation of the same algorithm. The benchmarks were run on a 2.0 GHz Intel Core i7 2630QM quad core CPU running 64-bit Windows 7 Home Premium SP1. The CPU code was compiled using the Microsoft C/C++ Optimizing Compiler version 16.00 and GPU benchmarks were performed using the NVIDIA CUDA programming toolkit version 4.1 running on a NVIDIA GPUs GeForce GT540M device.\\

As execution parameters were chosen combinations of:

\begin{itemize}\addtolength{\itemsep}{-0.5\baselineskip}
\item Max number of iterations: 1000, 2000, 4000, 10000, 20000 and 40000;
\item Order (max degree) of polynomial expansion: 1, 2, 4 and 8;
\end{itemize}

The other parameters remain unchanged for all tests:

\begin{itemize}\addtolength{\itemsep}{-0.5\baselineskip}
\item Random mode for initial population: normal distribution in [-1, +1];
\item Type of error function (fitness): Threshold Mean Square Error (TMSE);
\item TMSE threshold: 0.49;
\item Selection criterion: both \emph{ranking} and \emph{roulette};
\item Training error threshold: 0.001, used as a stopping criteria;
\item Crossover application probability rate: 0.9;
\item Mutation application probability rate: 0.2;
\item Number of tournament chromosomes at each selection stage: 4;
\item Elitism chromosomes at each evolution: 2.
\end{itemize}

The experiments were done by using three implementations of the GAME model:

\begin{itemize}\addtolength{\itemsep}{-0.5\baselineskip}
\item serial: the CPU-based original implementation (serial code);
\item Opt: an intermediate version, obtained during the APOD process application (optimized serial code);
\item GPU: the parallelized version, as obtained at the end of an entire APOD process application.
\end{itemize}

For the scope of the present experiment, we have preliminarily verified the perfect correspondence between CPU- and GPU-based implementations in terms of classification performance.
 In fact, the scientific results for the serial version have been already evaluated and documented in a recent paper \citep{brescia2012a}, where the serial version of GAME was originally included in a set of machine learning models, provided within our team and compared with the traditional analytical methods.\\
Referring to the best results for the three versions of GAME implementation, we obtained the percentages shown in Tab. \ref{comp}.

\begin{table*}
\centering
\resizebox{13cm}{!}{
\begin{tabular}{l|lrccccc}
\hline
type of experiment &	 missing features & figure of merit & serial & Opt & GPU \\
\hline
complete patterns  & --	
                  & $\textit{class.accuracy}$		        &    $82.1$ & $82.2$  & $81.9$  \\
 &                  & $\textit{completeness}$		        &    $73.3$ & $73.0$  & $72.9 $ \\
  &                 & $\textit{contamination}$		    &    $18.7$ & $18.5$  & $18.8$  \\
\hline
without feat. 11         & 11
                 & $\textit{class.accuracy}$		        &    $81.9$ & $82.1$  & $81.7$  \\
  &                 & $\textit{completeness}	$	        &    $79.3$ & $79.1$  & $78.9$  \\
   &                & $\textit{contamination}$		    &    $19.6$ & $19.5$  & $19.8$  \\
\hline
only optical &   8, 9, 10, 11
 	    & $\textit{class.accuracy}$		                &    $86.4$ & $86.3$  & $86.0$  \\
  &                 & $\textit{completeness}$		        &    $78.9$ & $78.6$  & $78.4 $ \\
   &                & $\textit{contamination}$		    &    $13.9$ & $13.7$  & $14.1$  \\
\hline
mixed      &  5, 8, 9, 10, 11
 	    & $\textit{class.accuracy}$		                &    $86.7$ & $86.9$  & $86.5 $ \\
  &                 & $\textit{completeness}$		        &   $ 81.5$ & $81.4 $ & $ 81.2 $ \\
   &                & $\textit{contamination}$		    &    $16.6$ & $16.2$  & $16.7$  \\
\hline
\end{tabular}
}
\caption{Summary of the performances (in percentage) of the three versions of the GAME classifier. There are reported results for the four main dataset feature pruning experiments, respectively with all 11 features, without the last structural parameter (tidal radius), with optical features only and the last one without 5 features (mixed between optical and structural types).}
\label{comp}
\end{table*}

The results were evaluated by means of three statistical figures of merit, for instance \emph{completeness, contamination and accuracy}. However, these terms are differently defined by astronomers and \emph{data miners}.
In this case, for \emph{classification accuracy} we intend the fraction of patterns (objects) which are correctly classified (either GCs or non-GCs) with respect to the total number of objects in the sample; the \emph{completeness} is the fraction of GCs which are correctly classified as such and finally, the \emph{contamination} is the fraction of non-GC objects which are erroneously classified as GCs. In terms of accuracy and completeness, by using all available features but the number 11 (the tidal radius), we obtain marginally better results, as can be expected given the high noise present in this last parameter, which is affected by the large background due to the host galaxy light. In terms of contamination, better results are obtained by removing structural parameters, demonstrating the relevance of information carried by optical features in the observed objects (in particular, the isophotal and aperture magnitudes and the FWHM of the image were recognized as the most relevant by all pruning tests). Moreover, these experiments have also the advantage to reduce the number of features within patterns, without affecting  the classification performance. The less numerous are the patterns, the shorter the execution time of the training phase, thus providing a benefit to the overall computational cost of the experiments. \\
Finally, it is worth pointing out that the performance results quoted in Tab. \ref{comp} are all referred to the test samples (without considering the training results), and do not include possible biases affecting the KB itself. Hence they rely on the assumption that the KB is a fair and complete representation the \emph{real} population that we want to identify.\\
From the computational point of view, as expected, for all the three versions of GAME, we obtained consistent results, slightly varying in terms of less significant digits, trivially motivated by the intrinsic randomness of the genetic algorithms and also by the precision imposed by different processing units. We recall in fact, that the GPU devices used for experiments are optimized for single precision and they may reveal a little lack of performance in the case of double precision calculations. GPU devices, certified for double precision would be made available for testing in the first quarter of 2013, when optimized \textit{Kepler} GPU class type devices will be commercially distributed.\\
After having verified the computing consistency among the three implementations, we investigated the analysis of performance in terms of execution speed.\\
We performed a series of tests on the parallelized version of the model in order to evaluate the scalability performance with respect to the data volume. These tests make use of the dataset used for the scientific experiments (see Sec. \ref{Astrodata}), extended by simply replicating its rows several times, thus obtaining a uniform group of datasets with incremental size.\\
By considering a GPU device with $96$ cores, $2GB$ of dedicated global memory and a theoretical throughput of about $14GB/sec$ on the PCI-E bus, connecting the CPU (hereinafter Host) and GPU (hereinafter Device), we compared the execution of a complete training phase, by varying the degree of the polynomial expansion (function \textit{polytrigo}) and the size of the input dataset.\\
Tab. \ref{benchgpu} reports the results, where the training cycles have been fixed to $4000$ iterations for simplicity. The derived quantities are the following:

\begin{itemize}\addtolength{\itemsep}{-0.5\baselineskip}
\item HtoD (Host to Device): time elapsed during the data transfer from Host to Device;
\item DtoH (Device to Host): the opposite of HtoD;
\item DtoD (Device to Device): time elapsed during the data transfers internally to the device global memory;
\item DP (Device Processing): processing time on the GPU side;
\item HP (Host Processing): processing time on the CPU side;
\item P (Processing): total duration of the training processing phase (excluding data transfer time);
\item T (Transfer): total duration of various data transfers between Host and Device;
\item TOT (Total): Total time duration of a training execution;
\end{itemize}

The relationships among these terms are the following:

\begin{equation}
P = DP + HP
\label{eq11}
\end{equation}
\begin{equation}
T = HtoD + DtoH + DtoD
\label{eq12}
\end{equation}
\begin{equation}
TOT = P + T
\label{eq13}
\end{equation}

The total execution time of a training phase, given by Eq. \ref{eq13}, can be obtained by adding the processing time (on both Host and Device), given by Eq. \ref{eq11}, to the data transfer time (Eq. \ref{eq12}). However, in principle, in Eq. \ref{eq12} it would be necessary to calculate also the time elapsed during data transfers through the \textit{shared} memory of the Device, i.e. the data flow among all threads of the GPU blocks. In our case this mechanism is automatically optimized by Thrust itself, which takes care of the flow, thus hiding the thread communication setup to the programmer.\\
By comparing the time spent to transfer data between Host and Device (sum of columns HtoD and DtoH) with the processing time (column P), it appears an expected not negligible contribution of transfer time, well known as one of the most significant bottlenecks for the current type of GPUs (i.e. before the Kepler technology). In this case, a particular effort is required in the code design to minimize as much as possible this data flow. In our case, as the data size increases, such contribution remains almost constant (approximately $9\%$), thus confirming the correctness of our approach in keeping this mechanism under control.\\
In terms of work distribution between Host and Device, by comparing the processing time between Host and Device (respectively, columns HP and DP), for all data size the percentage of GPU calculation time remains almost the same (approximately $70\%$ of the whole processing time, given in column P).
Furthermore, by evaluating the average time needed to process one MB of data at different polynomial degrees, when the size grows from $0.15$ to $512$ MB, we obtain on average $\sim0.058$ MB/sec with degree = $1$, $\sim0.055$ MB/sec with degree = $2$, $\sim0.046$ MB/sec with degree = $4$, and $\sim0.036$ MB/sec with degree = $8$. Of course, as it was expected, the increase in the polynomial degree affects the average time to process a single MB of data, but of an amount which is small in respect of the growing size of data.\\
In conclusions, by considering both scientific and computing performances of the GPU based model, we can conclude that a polynomial expansion of degree $8$ is a good compromise between the approximation of the desired training error and the processing time needed to reach it, still maintaining good performances also in the case of large volumes of training data.

\begin{table*}
\centering
\resizebox{13cm}{!}{
\begin{tabular}{l|l|rcccccccc}
\hline
degree            & $data size$ &            &            &           & $Time$    & $[sec]$  &            &           &          \\
\hline\hline
                  &             & $DP$       & $HP$       & $HtoD$    & $DtoH$    & $DtoD$   & $P$        & $T$       & $TOT$      \\
\hline
$1$               &	            &            &            &           &           &          &            &           &          \\
                  & $0.15MB$    & $9.59$     & $5.16$     & $0.11$    & $0.11$    & $0.03$   & $14.75$    & $0.25$    & $15.00$    \\
                  & $16MB$	    & $152.12$   & $65.20$    & $10.00$   & $10.00$   & $2.68$   & $217.32$   & $22.68$   & $240.00$   \\
                  & $128MB$	    & $1191.26$  & $510.54$   & $82.00$   & $81.20$   & $5.50$   & $1701.80$  & $168.70$  & $1870.50$  \\
                  & $512MB$	    & $4919.60$  & $2108.40$  & $322.00$  & $322.00$  & $8.00$   & $7028.00$  & $652.00$  & $7680.00$  \\
\hline
$2$               &	            &            &            &           &           &          &            &           &          \\
                  & $0.15MB$    & $10.38$    & $5.35$     & $0.12$    & $0.12$    & $0.04$   & $15.73$    & $0.27$    & $16.00$    \\
                  & $16MB$	    & $161.74$   & $69.32$    & $11.00$   & $11.00$   & $2.95$   & $231.05$   & $24.95$   & $256.00$   \\
                  & $128MB$	    & $1265.59$  & $542.39$   & $90.20$   & $89.32$   & $6.05$   & $1807.98$  & $185.57$  & $1993.55$  \\
                  & $512MB$	    & $5232.36$  & $2242.44$  & $354.20$  & $354.20$  & $8.80$   & $7474.80$  & $717.20$  & $8192.00$  \\
\hline
$4$               &	            &            &            &           &           &          &            &           &          \\
                  & $0.15MB$    & $12.72$    & $5.98$     & $0.13$    & $0.13$    & $0.04$   & $18.70$    & $0.30$    & $19.00$    \\
                  & $16MB$	    & $193.59$   & $82.97$    & $12.10$   & $12.10$   & $3.24$   & $276.56$   & $27.44$   & $304.00$   \\
                  & $128MB$	    & $1517.58$  & $650.39$   & $99.22$   & $98.25$   & $6.66$   & $2167.98$  & $204.13$  & $2372.11$  \\
                  & $512MB$	    & $6257.36$  & $2681.72$  & $389.62$  & $389.62$  & $9.68$   & $8939.08$  & $788.92$  & $9728.00$  \\
\hline
$8$               &	            &            &            &           &           &          &            &           &          \\
                  & $0.15MB$    & $16.33$    & $7.34$     & $0.14$    & $0.14$    & $0.04$   & $23.67$    & $0.33$    & $24.00$    \\
                  & $16MB$	    & $247.67$   & $106.14$   & $13.31$   & $13.31$   & $3.57$   & $353.81$   & $30.19$   & $384.00$   \\
                  & $128MB$	    & $1947.10$  & $834.47$   & $109.14$  & $108.08$  & $7.32$   & $2781.58$  & $224.54$  & $3006.12$  \\
                  & $512MB$	    & $7994.13$  & $3426.06$  & $428.58$  & $428.58$  & $10.65$  & $11420.19$ & $867.81$  & $12288.00$ \\
\hline
\end{tabular}
}
\caption{Summary of the time performances of the GPU version of the GAME classifier. The figures were obtained by varying the degree of the polynomial function \textit{polytrigo} (first column) and the size of input dataset (second column). The training cycles have been fixed to $4000$ iterations.
For each degree value, the first row (with datasize = $0.15MB$) is referred to the original size of the dataset used for scientific experiments. The other rows are referred to the same dataset but artificially extended by replicating its rows a number of times. All other columns 3-10 are time values expressed in seconds. Their meaning is defined as follows: $DP$ (Device Processing) is the processing time on the GPU side; $HP$ (Host Processing) is the processing time on the CPU side; $HtoD$ (Host to Device) is the time elapsed during the data transfer from Host to Device; $DtoH$ (Device to Host) is the opposite of $HtoD$; $DtoD$ (Device to Device) is the time elapsed during the data transfers internally to the device global memory; $P$ is the sum of columns $DP$ and $HP$; $T$ is the sum of columns $HtoD$, $DtoH$ and $DtoD$ and finally the last column $TOT$ is the sum of columns $P$ and $T$.}
\label{benchgpu}
\end{table*}

The three versions of GAME, respectively the serial one, the serial optimized and the parallel, have been tested under the same setup conditions. As expected, while the two CPU-based versions, serial and Opt, appear comparable, there is an evident difference with the GPU version. The diagrams shown in Fig.~\ref{figure4} report the direct comparisons among the three GAME versions, by setting an incremented degree of the polynomial expansion which represents the evaluation function for chromosomes.

\begin{figure}
\centering
\includegraphics[width=7.4cm]{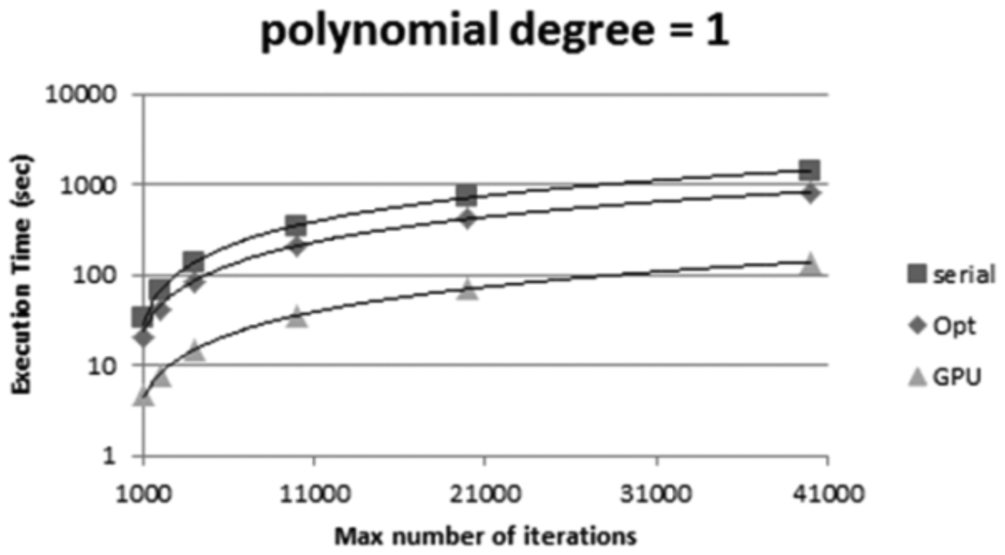}(a)
\includegraphics[width=7.4cm]{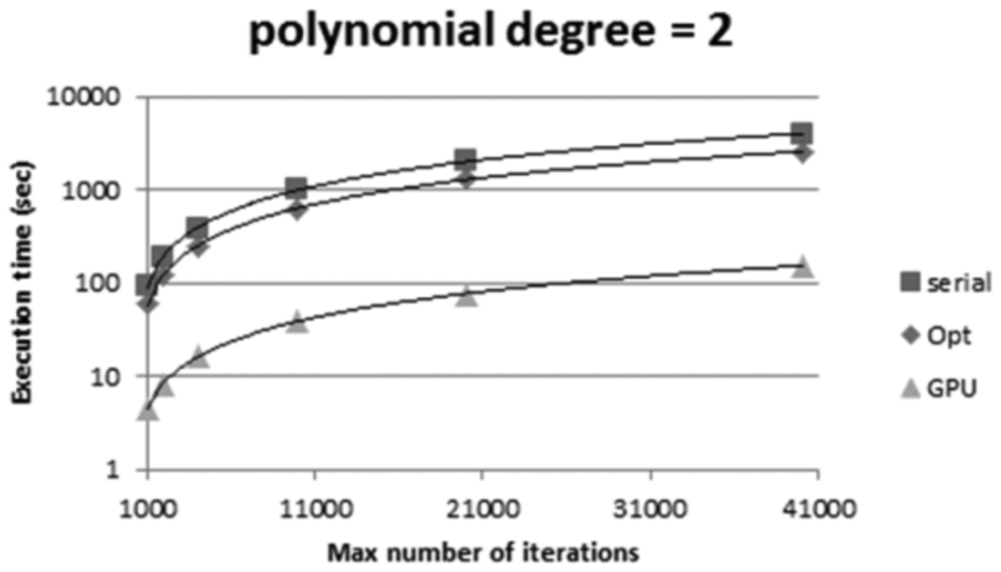}(b)
\includegraphics[width=7.4cm]{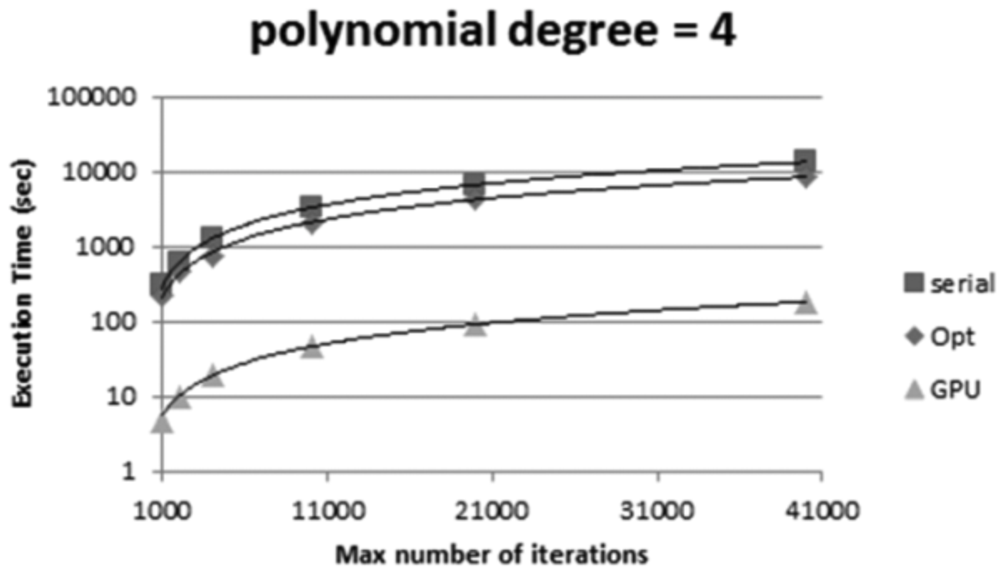}(c)
\includegraphics[width=7.4cm]{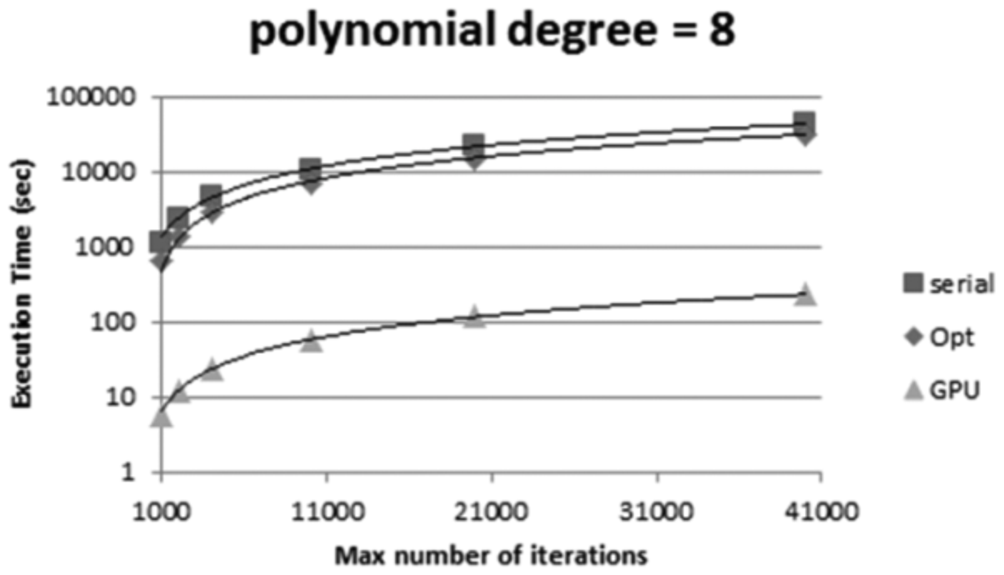}(d)
\caption{Comparison among the three GAME implementations with the polynomial degree 1 (a), 2 (b), 4 (c) and 8 (d). The squares represent the serial version; rhombi represent the Opt version, while triangles are used for the GPU version.}
\label{figure4}
\end{figure}

The trends show that the execution time increases always in a linear way with the number of iterations, once fixed the polynomial degree. This is what we expected because the algorithm repeats the same operations at each iteration. The GPU version speed is always at least one order of magnitude less than the other two implementations. We remark also that the classification performance of the GAME model increases by growing the polynomial degree, starting to reach good results from a value equal to 4. Exactly when the difference between CPU and GPU versions starts to be 2 orders of magnitude.

\begin{figure}
\centering
\includegraphics[width=7.5cm]{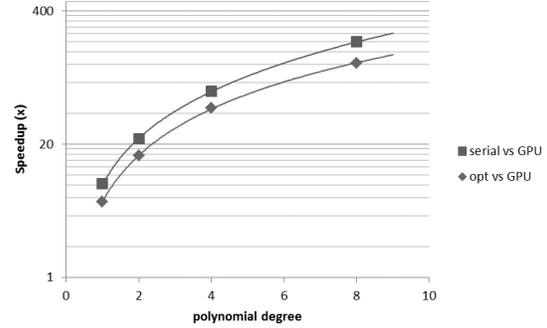}
\caption{Speedup comparison among GAME CPU implementations (serial and Opt) against the GPU version.}
\label{figure5}
\end{figure}

In the diagram of Fig.~\ref{figure5}, the GPU version is compared against the CPU implementations. As shown, the speedup increases proportionally with the increasing of the polynomial degree. The diagram shows that for the average speed in a range of iterations from 1000 to 40000, the GPU algorithm exploits the data parallelism as much data are simultaneously processed. As previously mentioned, an increase of maximum degree in the polynomial expansion leads to an increase in the number of genes and consequently to a larger population matrix. The GPU algorithm outperforms the CPU performance by a factor ranging from $8\times$ to $200\times$ against the serial version and in a range from $6\times$ to $125\times$ against the Opt version, enabling an intensive and highly scalable use of the algorithm that were previously impossible to be achieved with a CPU.

\section{Conclusions}

A multi-purpose genetic algorithm implemented with GPGPU/CUDA parallel computing technology has been designed and developed. The model comes from the paradigm of supervised machine learning, addressing both the problems of classification and regression applied on medium/massive datasets.\\
The GPU version of the model is foreseen to be deployed on the DAMEWARE web application, presented in \citet{brescia2010} and \citet{brescia2011}. The model has been successfully tested and validated on astrophysical problems \citep{brescia2012a}.\\
Since genetic algorithms are inherently parallel, the parallel computing paradigm provided an exploit of the internal training features of the model, permitting a strong optimization in terms of processing performance. In practice, the usage of CUDA translated into a $75\times$ average speedup, by successfully eliminating the largest bottleneck in the multi-core CPU code. Although a speedup of up to $200\times$ over a modern CPU is impressive, it ignores the larger picture of using a Genetic Algorithm as a whole.\\

Real datasets can be very large (those we have previously called Massive datasets) and this requires greater attention to GPU memory management, in terms of scheduling and data transfers host-to-device and \emph{vice versa}.

We wish also to remark that the fact that the three different implementations (serial, serial optimized and GPU)  of the GAME method lead to identical scientific results, implies that the three versions are fully consistent. Furthermore, by considering the good results of a deep analysis of the processing time profile, as expected, the GPU version largely improves the scalability of the method in respect of massive data sets.\\
Finally, the very encouraging results suggest to investigate further optimizations, like: (i) moving the formation of the population matrix and its evolution in place on the GPU. This approach has the potential to significantly reduce the number of operations in the core computation, but at the cost of higher memory usage; (ii) exploring more improvements by mixing Thrust and CUDA C code, that should allow a modest speedup justifying development efforts at a lower level; (iii) use of new features now available on NVIDIA Kepler architecture, such as OpenACC directives, able to achieve faster atomics and more robust thread synchronization and multi GPUs capability.

\subsubsection*{Acknowledgments}

This work originated from a M.Sc. degree  in Informatics Engineering done in the context of a collaboration among several Italian academic institutions. The hardware resources were provided by Dept. of Computing Engineering and Systems and S.Co.P.E. GRID Project infrastructure of the University Federico II of Naples. The data mining model has been designed and developed by DAME Program Collaboration.
\noindent MB wishes to thank the financial support of PRIN-INAF 2010, \emph{Architecture and Tomography of Galaxy Clusters}. This work has been partially funded by LINCE project of the F.A.R.O. programme jointly financed by the Compagnia di San Paolo and by the Polo delle Scienze e delle Tecnologie of the University Federico II of Napoli and it has been carried out also thanks to a hardware donation in the context of the NVIDIA Academic Partnership program. Finally, a special thanks goes to the anonymous referee for all very useful comments and suggestions.

\bibliographystyle{model2-names}



\end{document}